\documentclass{mn2e}

\input{epsf}

\def\eg{{\rm e.g.\ }}

\def\etal{{\rm et al.\ }}

\def\gsim{\ga}
\def\lsim{\la}
\def\simprop{ \lower .75ex \hbox{$\sim$} \llap{\raise .27ex \hbox{$\propto$}} }

\def\Msol{M_\odot}

\def\kms{{\rm km\,s^{-1}}}

\def\mum{\mu{\rm m}}
\def\Myr {{\rm Myr}}
\def\Gyr {{\rm Gyr}}
\def\yr {{\rm yr}}
\def\mjy {{\rm \,mJy}}

\def\tesc {t_{\rm esc}}

\begin{document}


\title[Can the faint sub-mm galaxies be explained in the $\Lambda$CDM model?]
{Can the faint sub-mm galaxies be explained in the $\Lambda$CDM model?}


\author[Baugh \etal]{
\parbox[t]{\textwidth}{
\vspace{-1.0cm}
C. M. Baugh$^{1}$,
C. G. Lacey$^{1}$,
C. S. Frenk$^{1}$,
G. L. Granato$^{2}$,
L. Silva$^{3}$,
A. Bressan$^{2}$,
A. J. Benson$^{4}$,
S. Cole$^{1}$.
}
\vspace*{6pt} \\ 
$^{1}$Institute for Computational Cosmology, Department of Physics, 
University of Durham, South Road, Durham DH1 3LE, UK.\\ 
$^{2}$Osservatorio Astronomico di Padova, Vicolo dell'Osservatorio, 5, I-35122 Padova, Italy.\\
$^{3}$Osservatorio Astronomico di Trieste, via Tiepolo 11, I34131 Trieste, Italy.\\
$^{4}$Astrophysics, University of Oxford, Keble Road, Oxford, OX1 3RH, UK. \\
\vspace*{-0.5cm}}

\maketitle 
 
\begin{abstract}
We present predictions for the abundance of sub-mm galaxies (SMGs) and
Lyman-break galaxies (LBGs) in the $\Lambda$CDM cosmology.  A key
feature of our model is the self-consistent calculation of the
absorption and emission of radiation by dust.  The new model
successfully matches the LBG luminosity function, as well as reproducing
the properties of the local galaxy population in the optical and IR.
The model can also explain the observed galaxy number counts at
$850\mum$, but only if we assume a top-heavy IMF for the stars formed
in bursts.  The predicted redshift distribution of SMGs depends
relatively little on their flux over the range $1$-$10\mjy$, with a
median value of $z\approx 2.0$ at a flux of $5\mjy$, in good
agreement with the recent measurement by Chapman \etal The counts of
SMGs are predicted to be dominated by ongoing starbursts.  However, in
the model these bursts are responsible for making only a few per cent
of the stellar mass locked up in massive ellipticals at the present
day.
\end{abstract}


\section{Introduction} 
\label{sec:intro}

The detection of populations of star-forming galaxies at high redshift
($z\gsim 2)$ in the late 1990s opened a window on the process of
galaxy formation when the Universe was less than 20\% of its current
age. Two main techniques have been used to discover such objects: (1)
Lyman-break galaxies (LBGs) are detected in optical bands from their
stellar emission in the rest-frame ultraviolet, using the spectral
break around 912\AA\ produced by absorption by intervening neutral
hydrogen (Steidel \etal 1996). The largest samples of LBGs have been
obtained for $z\sim 3-4$ (Steidel \etal 1999).  The dust extinction
corrections required to derive star formation rates of LBGs from their
rest-frame UV luminosities are probably quite large (a factor of $\sim 3 -
10$) (\eg Meurer \etal 1999; Adelberger \& Steidel 2000), but remain
uncertain, this being a major obstacle preventing an accurate
determination of the contribution of LBGs to the global star formation
rate. (2) Sub-mm galaxies (SMGs) are detected by emission from warm
dust in the rest-frame far-IR/sub-mm (Smail \etal 1997; Hughes \etal
1998). From their discovery using the SCUBA instrument on JCMT, the
faint $\mjy$-flux SMGs were suspected to lie at high redshifts 
(e.g. Yun \& Carilli 2002) 
and
this has now been firmly established by measurements of spectroscopic
redshifts, which give a median redshift $z\approx 2.2-2.4$ for sources
with $850\mum$ fluxes around 5\mjy\ (Chapman \etal 2003, 2004). The SMGs are
generally interpreted as dust-enshrouded starbursts, in which the
dust is being heated by UV radiation from young stars, but estimates of
their star formation rates, based on the measured $850\mum$ fluxes, are
uncertain because of a lack of knowledge of the shapes of their
spectral energy distributions (SEDs) at the shorter wavelengths which
presumably dominate the bolometric luminosity of the dust. There also
remains the possibility that a significant fraction of the sub-mm
emission may be powered by heating of dust by AGNs rather than young
stars. Recent X-ray observations suggest that the AGN contribution to
the sub-mm emission is small in most SMGs (Alexander \etal 2003), but
this issue will only be finally settled by further observational
data. 

A third technique that has been used to search for high-redshift
star-forming galaxies is to look for the Ly$\alpha$ emission line
(e.g. Hu, Cowie \& McMahon 1998). The usefulness of Ly$\alpha$
emission as a quantitative star formation indicator is, however, limited
by the fact that many star-forming galaxies do not show Ly$\alpha$ in
emission (e.g. Steidel \etal 2000), and that resonant scattering of
Ly$\alpha$ photons by neutral hydrogen makes corrections for dust
extinction even more uncertain than for the UV continuum. Thus, more
reliable estimates of the star formation rates in these systems
typically use measurements of the rest-frame UV luminosity, as in
LBGs. We present predictions from our semi-analytical model for
Ly$\alpha$ emitting galaxies in a separate paper (Le~Delliou \etal
2004), but do not consider them further here.

Within the uncertainties, galaxies detected as LBGs and as SMGs appear
to contribute comparably to the global star formation rate at high
redshifts (\eg Smail \etal 2002). In combination, they should, in
principle, provide a census of all high-mass star formation at
redshifts $z\gsim 2$, since one should either see the UV radiation
from these stars directly (as in LBGs) or after reprocessing by dust
(as in SMGs). It is therefore of fundamental importance to understand
both types of galaxy, in order to understand the process of galaxy
formation as a whole. In this paper, we describe the main results of
our attempt to do this within the framework of the CDM model of
structure formation; subsequent papers will describe different aspects
of our model in more detail.

We build on our previous work using semi-analytical models of galaxy
formation to calculate the evolution of the galaxy population {\em ab
initio}. These models predict the mass accretion, merging and star
formation history of each galaxy, along with the evolution of its gas
mass and metallicity, and the mass and size of the disk and bulge. In
Baugh \etal (1998) we interpreted the first observational results on
LBGs in terms of a hierarchical model of galaxy formation, but without
including any treatment of dust. In Granato \etal (2000, hereafter
G00), we presented a new, {\em multi-wavelength} version of our model
which included a comprehensive treatment of the reprocessing of
stellar radiation by dust (using the {\tt GRASIL} code developed by
Silva \etal 1998, hereafter S98), and showed that it reproduced the
luminosity functions and other properties of galaxies at $z=0$ all the
way from the far-UV to the far-IR. However, as we show in this paper,
when the G00 model in its original form is applied to the high-z
universe, it dramatically underpredicts both the number density of
bright LBGs and the number counts of SMGs at $850\mum$. Of these, the
latter problem seems the most serious, since the observed sub-mm
fluxes of SMGs are usually interpreted as requiring objects with star
formation rates in excess of $100 \Msol \yr^{-1}$ (for a standard
initial mass function) to be commonplace; such vigorous star formation
rates at high redshifts are difficult to accommodate in the
CDM model. To see if it is possible to achieve such high sub-mm
luminosities in high-z galaxies, we have therefore made significant
modifications to our original model, in regard to both star formation
and the behaviour in bursts, in order to arrive at a new
self-consistent model, which {\em at the same time} reproduces the
properties of the LBGs and SMGs at high redshift {\em and} the main
properties of the galaxy population at low redshift. We present here
the first results from this new model.

The challenge of modelling galaxies in the high redshift universe 
has received much attention in the past decade. Simple theoretical 
and empirical models 
have been presented which use the number counts of galaxies observed at 
different wavelengths to try to constrain the global star formation history 
of the Universe (e.g. Pearson \& Rowan-Robinson 1996; Blain \etal 1999a, 
1999b; Chary \& Elbaz 2001; Balland, Devriendt \& Silk 2003). More 
sophisticated models that attempt to follow the physics of galaxy 
formation in greater detail have stressed the importance of bursts 
of star formation at high redshift in explaining the abundance of SMGs 
(Guiderdoni \etal 1998; Devriendt \& Guiderdoni 2000) and LBGs 
(Somerville, Primack \& Faber 2001). 
Most recently, Granato \etal (2004) have developed a coupled model 
for the evolution of galactic spheroids and QSOs, which successfully 
reproduces the counts of SMGs and the evolution of the K-band 
luminosity function with redshift 
(e.g. Pozzetti \etal 2003; Drory \etal 2004).

Our methodology represents a number of significant advances over
earlier attempts to model the SMGs and LBGs in the framework of
hierarchical galaxy formation. Previous work has tried to model either
SMGs (\eg Guiderdoni \etal 1998; Blain \etal 1999b) 
or LBGs (\eg Baugh \etal 1998;
Somerville, Primack \& Faber 2001), but not both populations
together. This allowed them to tune their model parameters to fit
either one population or the other. In this paper, we try to explain
both populations together. Our model has the following important
features, which improve on earlier work: (i) It directly predicts the
merger histories of dark matter haloes and of the galaxies contained
within them.  This allows us to predict the redshifts and gas masses
of bursts of star formation triggered by galaxy mergers.  (ii) The
absorption of starlight by dust is calculated self-consistently by
radiative transfer, using the dust mass obtained from the gas mass and
metallicity, and the disk and bulge scalelengths predicted by the
model.  This is particularly important when computing the amount of
extinction at short wavelengths. In turn, this determines the amount
of energy absorbed by the dust and re-radiated at long wavelenths.
(iii) The spectrum of dust emission is calculated using a distribution
of dust temperatures within each galaxy, which is obtained by solving
for the radiative equilibrium temperatures of individual dust grains.

In contrast to our detailed treatment of dust emission, previous works
either treated the dust temperature as a free parameter in fitting the
SMG number counts (\eg Kaviani, Haehnelt \& Kauffmann 2003), or used
empirical IR/sub-mm SEDs estimated from present-day galaxies (\eg
Pearson \& Rowan-Robinson 1996; Blain \etal 1999a; Devriendt \& Guiderdoni 2000). 
The task of fitting the observed number
counts of SMGs is much simpler if the dust temperature is treated as a
free parameter: for a dust emissivity $\epsilon_{\nu} \propto
\nu^{\beta}$, bolometric dust luminosity $L_d$, and temperature $T_d$,
the luminosity per unit frequency at long wavelengths scales as
$L_{\nu} \propto \nu^{2+\beta} L_d/T_d^{3+\beta}$, which varies
roughly as $T_d^{-5}$ at fixed $L_d$ for $\beta\approx 2$. This might
suggest that it is easy to reconcile any theoretical model with the
observed sub-mm counts simply by making modest adjustments to the
assumed dust temperature. However, treating the dust temperature as a
free parameter is physically inconsistent, since it is actually
determined by the condition that the dust grains be in thermal
equilibrium between radiative heating and cooling. For dust mass
$M_d$, with a single dust temperature, thermal equilibrium implies
$L_d \propto M_d T_d^{\beta+4}$, so that $L_{\nu} \propto
\nu^{2+\beta} M_d^{(3+\beta)/(4+\beta)} L_d^{1/(4+\beta)}$ if $M_d$
and $L_d$ are treated as the physical parameters, rather than $T_d$
and $L_d$. This argument shows that once one imposes the condition
that dust temperatures be physically consistent, the predicted sub-mm
luminosities in a model can only be changed by large factors by making
large changes to the dust masses.

Additional shortcomings of previous works are that they either did not
include halo and galaxy mergers (\eg Guiderdoni \etal 1998), or
treated the amount of dust extinction in high-z galaxies as a free
parameter (\eg Somerville \etal 2001).

In the next Section we explain the major changes to the model since 
G00. We present the results from our model in Section~\ref{s:results},
and our conclusions  in Section~\ref{s:conc}.

\section{The model}

We now give a brief overview of our new galaxy formation model.  In
Section \ref{ssec:basic}, for completeness, we list some general
differences from the model presented by G00, which were motivated by
changes in the preferred values of the cosmological parameters and by
improvements in the modelling of feedback processes (Benson \etal
2003; hereafter B03).  We demonstrate in Section \ref{s:results} that
neither the G00 model nor the model presented by B03 can reproduce the
number counts of sub-mm galaxies. The additional changes required to
the model of B03 to rectify this are set out in detail in Section
\ref{ssec:advanced}.  The impact of these changes on the predicted
sub-mm counts is demonstrated in Fig. \ref{fig:changes} in Section
\ref{s:results}.  Full details of the semi-analytic galaxy formation
model can be found in Cole \etal (2000) and B03. The {\tt GRASIL}
model for the emission from stars and dust is described by S98, and
its implementation in the semi-analytic model is described in G00.
A full description of our new model will be given in Lacey \etal
(2004a).

\subsection{The basic model}
\label{ssec:basic}

The galaxy formation model from which we start is the one described by B03. 
This contains a number of changes from the model of G00. We assume a 
flat, $\Lambda$CDM cosmology, with density parameter $\Omega_{0}=0.3$, 
a Hubble constant of $h=H_{0}/(100 {\rm kms}^{-1}{\rm Mpc}=0.7$ and 
with a perturbation amplitude set by the linear {\it rms} fluctuation 
in spheres of radius $8h^{-1}$Mpc of $\sigma_{8}=0.9$.

One of the major differences is motivated by a revision in the value
of the cosmological baryon density, $\Omega_{b}$, to be consistent
with recent constraints from Big Bang Nucleosynthesis models (O'Meara
\etal 2001) and from measurements of fluctuations in the cosmic
microwave background radiation (e.g. Spergel \etal 2003).  We adopt a
value of $\Omega_{b}=0.04$, which is twice that used by G00. To
prevent the formation of too many luminous galaxies, B03 found it
necessary to invoke a ``superwind'' mode of feedback, in which cold
gas is ejected from a galaxy in proportion to the star formation
rate. The superwind operates in addition to the feedback used by G00
(see Cole \etal 2000).  There is observational support for the
existence of superwinds from the detection of high-velocity outflows
in massive galaxies at both low and high redshift (\eg 
Martin 1999; Heckman \etal 2000; 
Pettini \etal 2002; Adelberger \etal 2003; 
Smail \etal 2003).

The B03 model also includes a simple treatment of the feedback from
photoionization of the intergalactic medium, which suppresses the
collapse of gas into low mass dark halos after reionization. We model
this by assuming that no gas cools in halos with circular velocities
$V_c < 60{\rm kms}^{-1}$ at $z<6$. This has a similar effect on the
predicted form of the galaxy luminosity function to the more realistic
and sophisticated treatment developed by Benson \etal (2002b).

The {\tt GRASIL} code computes the emission from both the stars and dust in
a galaxy, based on the star formation and metal enrichment histories
predicted by the semi-analytical model. It includes radiative transfer
through a two-phase dust medium, with a diffuse component and giant
molecular clouds, and a distribution of dust grain sizes. Stars are
assumed to form inside the clouds and then gradually leak out. The
output from {\tt GRASIL} is the galaxy SED from the far-UV to the sub-mm.
The motivation for the choice of dust parameters in the {\tt GRASIL} model 
and the impact of varying these choices are discussed in detail in 
G00 (see also Prouton \etal 2004). 

\subsection{Key features of the new model}
\label{ssec:advanced}

In Section \ref{s:results}, we will show that the predictions of the
G00 and B03 models do not match the observed number counts of sub-mm
sources.  The positive effect on the number counts of increasing the
baryon fraction by a factor of two between G00 and B03 is offset by
the inclusion of superwinds in the latter.  It is therefore necessary
to make a number of important changes to the B03 model in order to
reproduce the SMG counts.  Our freedom to change the B03 model is,
however, restricted for two reasons.  Firstly, we retain the
cosmological parameters adopted by B03. This sets the rate at which
structure grows in the dark matter, which in turn determines the halo
merger trees in which galaxies form and evolve.  Secondly, we follow
the philosophy set out by Cole \etal (2000), and demand that any new
model should reproduce basic properties of the local galaxy
population, such as the luminosity functions in the optical, near-IR
and far-IR, the distribution of disk scale sizes, the gas and metal
contents of disks and spheroids, and the mix of morphological types.
The task of finding a successful model is made even more demanding
by our use of a self-consistent calculation of the absorption and
re-emission of radiation by dust. We
emphasize again that the temperature distribution of the dust grains
is not a free parameter in our model, but is determined by the stellar
luminosity and the mass, distribution and composition of the dust. We
largely retain the same parameters in the {\tt GRASIL} code as used by
G00, with modifications to the escape time of young stars from dust
clouds and to the long-wavelength emissivity of the dust grains, which
we discuss in the next sections.

The model set out below was reached after an extensive exploration of
parameter space, within the boundaries set by the stringent
constraints outlined above, with the goal of improving the model
predictions for the high redshift galaxy population.  The consequences
of systematically varying key model assumptions will be illustrated in
Section \ref{s:results}.

{\it Star formation timescale in disks:} 
The star formation rate is
given by $\psi = M_{\rm gas}/\tau_{\star }$, where $M_{\rm gas}$ is
the cold gas mass. G00 and B03 assumed that the star formation
timescale in disks, $\tau_{\star }$, incorporated a scaling with the
dynamical time of the disk. In this case, the star formation timescale
becomes shorter with increasing redshift. Consequently, mergers at
high redshift tend to be between gas-poor disks, with little fuel
available to power a starburst.  In order to move more star formation
from the quiescent to the burst mode at high redshifts, 
we adopt a prescription for the
star formation timescale in disks {\it without} the dynamical time scaling:
\begin{equation} 
\tau_{\star } = \tau_{\star 0}\left(V_{c}/200\kms \right)^{\alpha_{\star }},
\end{equation} 
where $\tau_{\star 0}$ is a constant. This parametrization of the star
formation timescale is similar to that employed in our earlier models
(Cole \etal 1994; Baugh \etal 1996, 1998).  There are no compelling
theoretical arguments for preferring one of these prescriptions over
the other. (We test the star formation recipe in our models against 
observational data in Bell \etal 2003.) 
In both cases, appropriate parameter values can be chosen
to reproduce the observed gas fraction-luminosity relation at $z=0$
(see Fig. 9 of Cole \etal 2000); here, we choose $\tau_{\star
0}=8\Gyr$ and $\alpha_{\star }=-3$.  In the new model, galactic disks
at low redshift have roughly the same star formation timescales as in
G00, but disks at high redshifts have much longer star formation
timescales than before.  As a consequence, in the new model disks at
high redshift are much more gas-rich than in G00, resulting in more
gas being available for star formation in starbursts triggered by mergers
between these disks. This type of picture is hinted at observationally
by the very large gas fractions implied by CO measurements of SMGs
(\eg Neri \etal 2003). Similar schemes resulting in gas-rich mergers
at high redshift have previously been invoked to explain the
luminosity function of Lyman-break galaxies (Somerville \etal 2001)
and the evolution of the quasar population (Kauffmann \& Haehnelt
2000).

{\it Star formation bursts triggered by galaxy mergers:} Bursts of
star formation are assumed to be triggered only by galaxy mergers. We
define a galaxy merger as {\em major} or {\em minor} according to
whether the ratio of masses of the two galaxies exceeds $f_{\rm
ellip}$ or not. In {\em major} mergers, the stellar disks are assumed
to be transformed into a new stellar spheroid, while in {\em minor}
mergers, the stellar disk of the larger galaxy is preserved. Bursts are
assumed to be triggered in all major mergers (as in G00), but also
(unlike in G00) in minor mergers which satisfy both of the following
conditions: (i) the galaxy mass ratio exceeds $f_{\rm burst}$ (where
$f_{\rm burst} < f_{\rm ellip}$); (ii) the gas fraction in the primary
galaxy
exceeds $f_{\rm gas,crit}$.  The latter condition is motivated by the
suggestion of Hernquist \& Mihos (1995) that gas-rich disks should be
more susceptible to bursts triggered by the accretion of small
satellites. We adopt values of $f_{\rm ellip}=0.3$, $f_{\rm
burst}=0.05$ and $f_{\rm gas,crit}=0.75$.  Another modification is
that the timescale of star formation in a burst is taken to be
$\tau_{\star \rm burst} = {\rm max}\,[f_{\rm dyn} \tau_{\rm dyn},
\tau_{\star \rm burst,min}]$, where $\tau_{\rm dyn}$ is the dynamical
time of the newly formed spheroid (G00 assumed $\tau_{\star \rm burst}
= f_{\rm dyn} \tau_{\rm dyn}$). We adopt $f_{\rm dyn}=50$ and
$\tau_{\star \rm burst,min}=0.2\Gyr$; these values were found to produce 
the best match to the present day $60\mu m$ luminosity function and the 
abundance of SMGs.

{\it The IMF in bursts:} Perhaps the most dramatic change from our
previous work is the adoption of a top-heavy initial mass function
(IMF) for stars formed in bursts. This is the single most important
change to the model, which has the biggest impact on the predicted
counts of sub-mm sources.  For stars formed quiescently in disks, we
use a solar neighbourhood IMF, with the form proposed by Kennicutt
(1983): ${\rm d}N/{\rm d ln }m \propto m^{-x}$, with $x=0.4$ for
$m<1M_{\odot}$ and $x=1.5$ for $m>1M_{\odot}$. In bursts, we instead
use a single power law with $x=0$. In both cases, the IMF covers a
mass range $0.15 < m < 125 M_{\odot}$. (In G00, we used a Kennicutt
IMF for all star formation.) There are two factors motivating the
adoption of a top-heavy IMF in starbursts.  Firstly, the total energy
radiated in the UV per unit mass of stars formed is increased due to
the larger fraction of high mass stars, which increases the amount of
radiation heating the dust.  For example, the energy per unit mass
emitted at $1500$\AA\ is around 4 times larger for the top-heavy $x=0$
IMF compared to the Kennicutt IMF.  Secondly, a top-heavy IMF also
results in a higher yield of metals from Type~II supernovae, so that
more dust is produced to absorb this energy.  For stars formed with
the top-heavy IMF, more than twice the mass is recycled to the
interstellar medium and over six times as many metals are produced,
compared with the case of stars formed with a Kennicutt IMF. The
increased production of dust for a top-heavy IMF is essential for
boosting the luminosity of galaxies in the sub-mm range, as discussed
in the Introduction. If the stellar UV luminosity were to be increased
without changing the mass of dust, the dust would be heated to higher
temperatures, shifting the peak of the dust emission spectrum to
shorter wavelengths, and resulting in relatively little increase in
the luminosity at wavelengths longwards of the peak (including the
sub-mm). Including the increased dust production, using a top-heavy
IMF results in the most efficient production of luminosity at sub-mm
wavelengths for a given rate of star formation.

\section{Results}
\label{s:results}

\begin{figure}
{\epsfxsize=8.5truecm
\epsfbox[0 150 590 700]{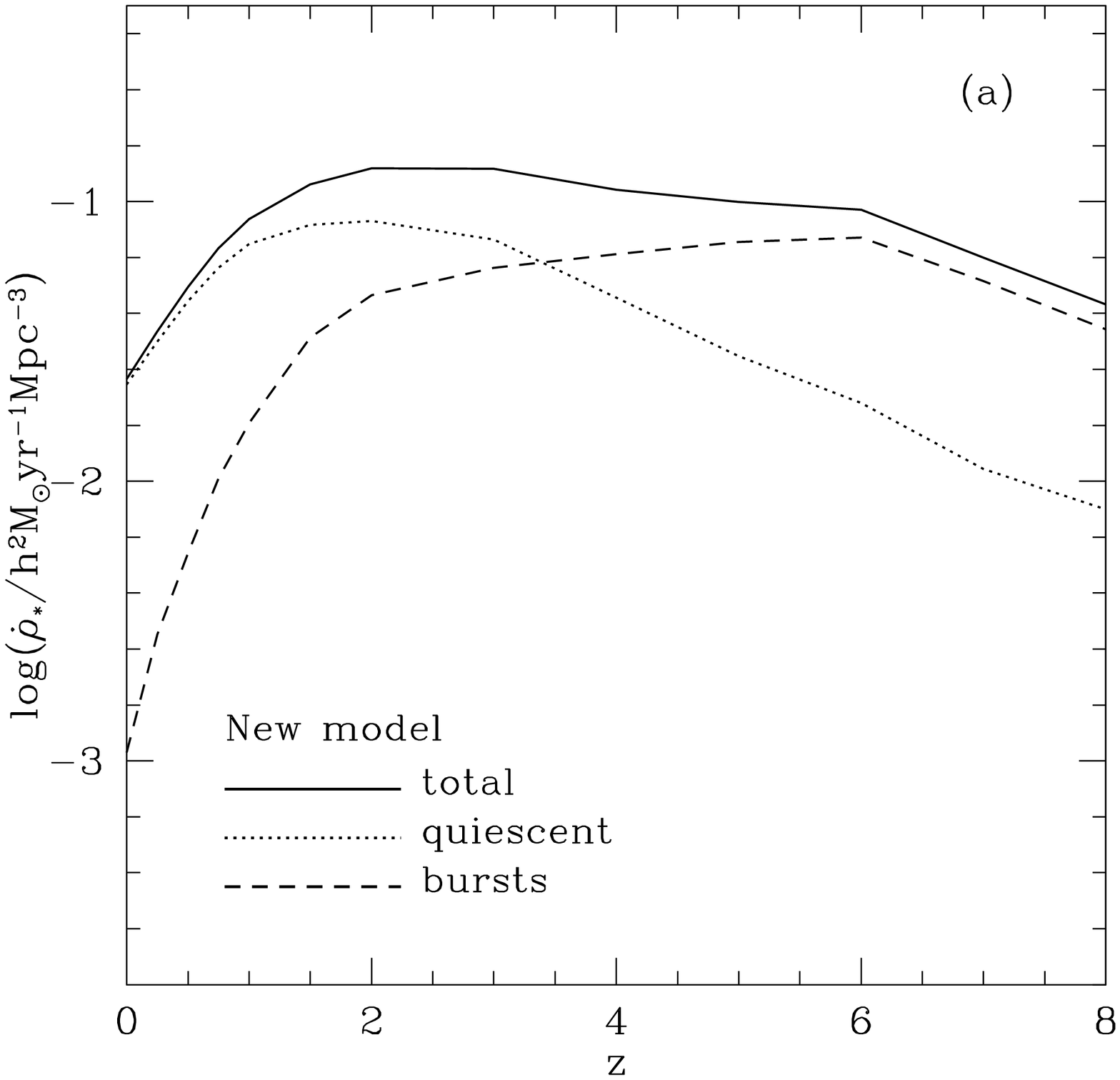}}
{\epsfxsize=8.5truecm
\epsfbox[0 150 590 700]{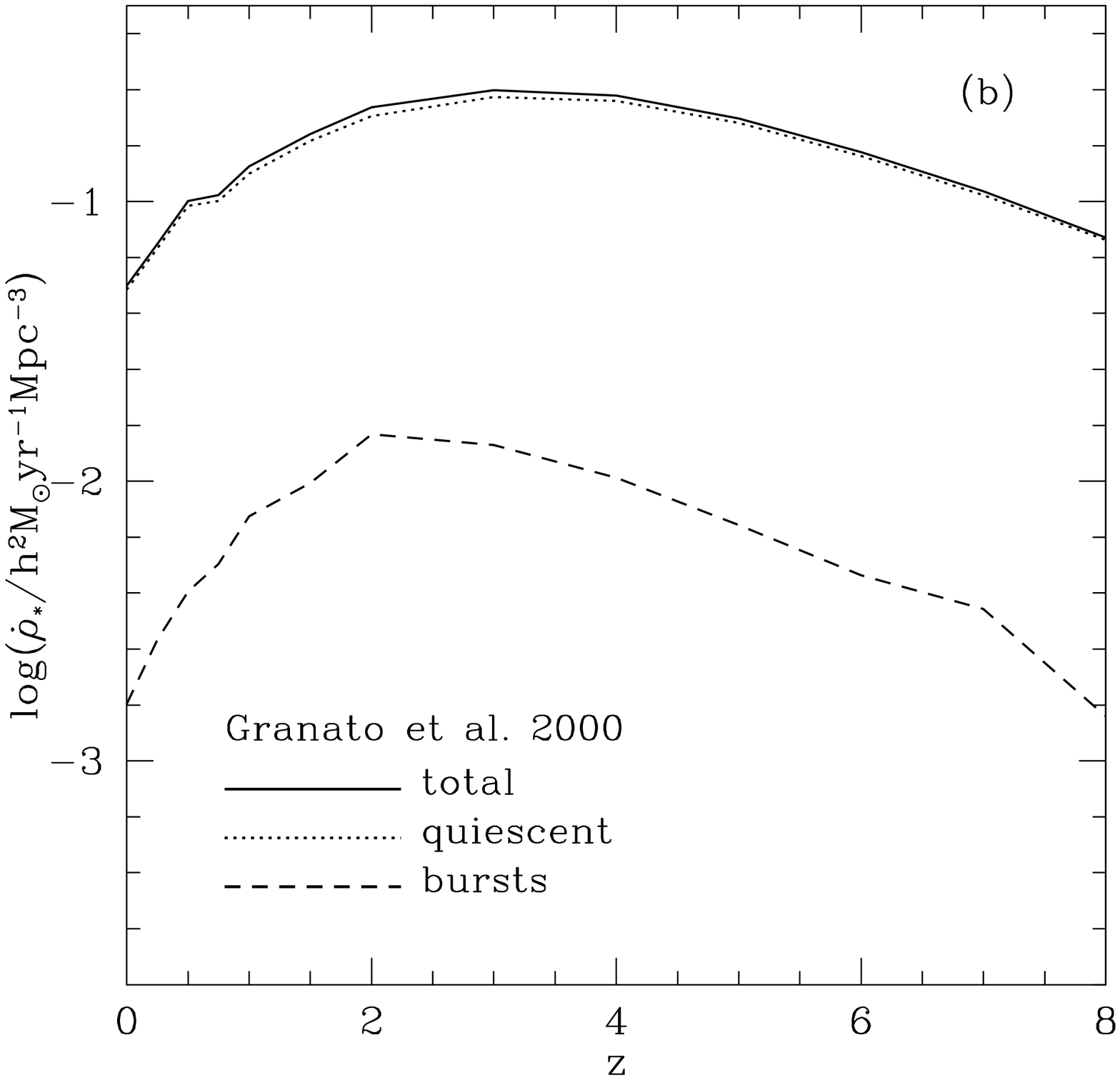}}
\caption{ The visible star formation rate (i.e. excluding brown
dwarfs) per unit volume as a function of redshift. (a) The predictions
of the new model presented in this paper. (b) The equivalent results for
the G00 model.  In both cases, the solid line shows the total star
formation rate density, the dotted line indicates the quiescent star
formation in galactic disks, and the dashed line shows the star
formation in bursts triggered by galaxy mergers.  }
\label{fig:sfrv}
\end{figure}

\begin{figure}
{\epsfxsize=8.5truecm
\epsfbox[0 150 590 700]{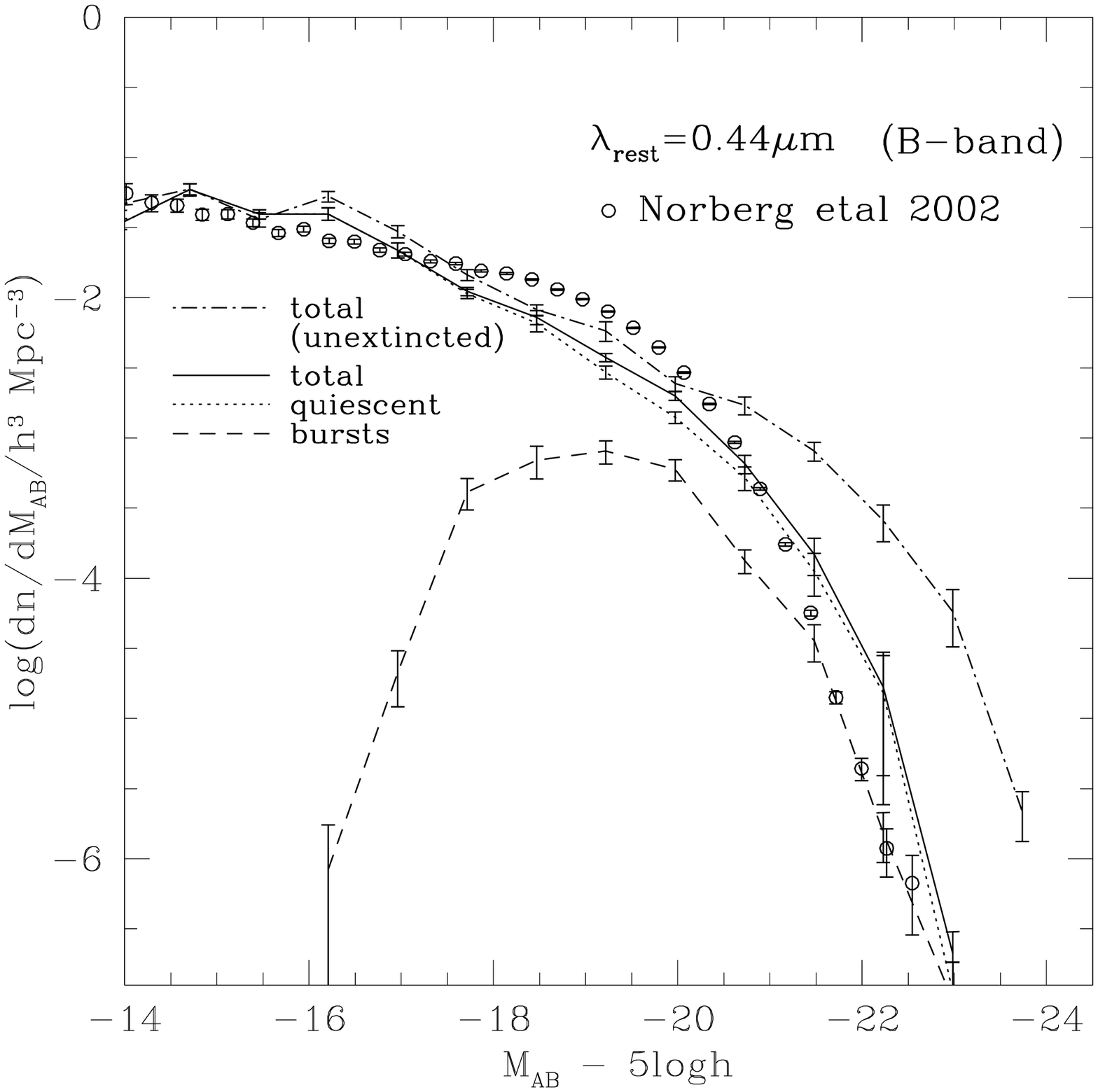}}
{\epsfxsize=8.5truecm
\epsfbox[0 150 590 700]{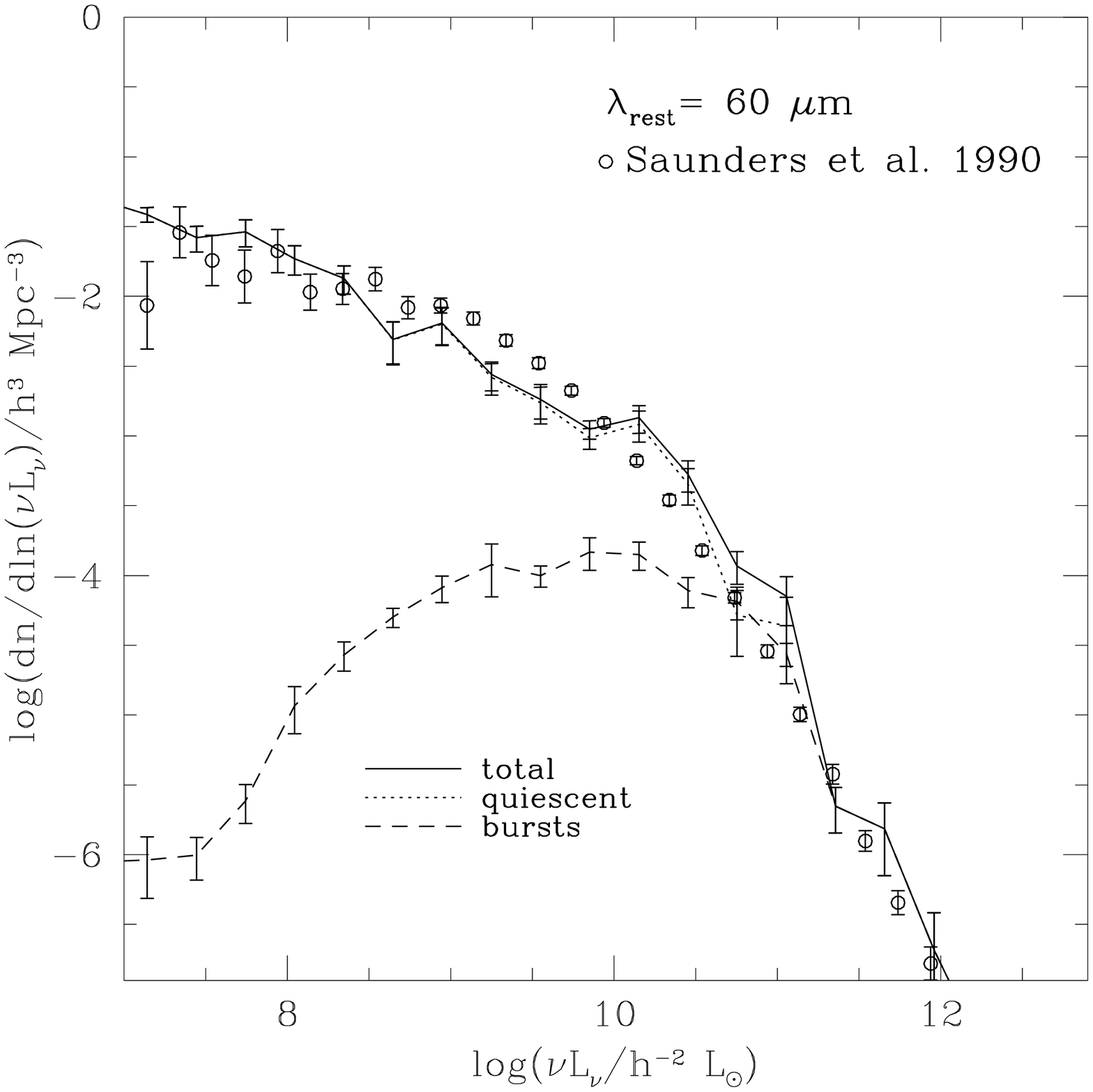}}
\caption{ The predictions for the optical (top) and far-IR (bottom)
luminosity functions at $z=0$ for the new model.  The dotted lines
show the luminosity function of galaxies that are forming stars
quiescently, the dashed lines show the contribution from ongoing
bursts, and the solid lines show the total luminosity function.  In the
upper panel, the dot-dashed line shows the predicted luminosity
function in the absence of dust extinction.  The error bars on the
model predictions indicate the Poisson error arising from the number
of galaxies simulated. The open symbols with error bars show the
observational data, taken from the sources given in the legend.}
\label{fig:lfz0}
\end{figure}

\begin{figure}
{\epsfxsize=8.5truecm
\epsfbox[60 161 377 700]{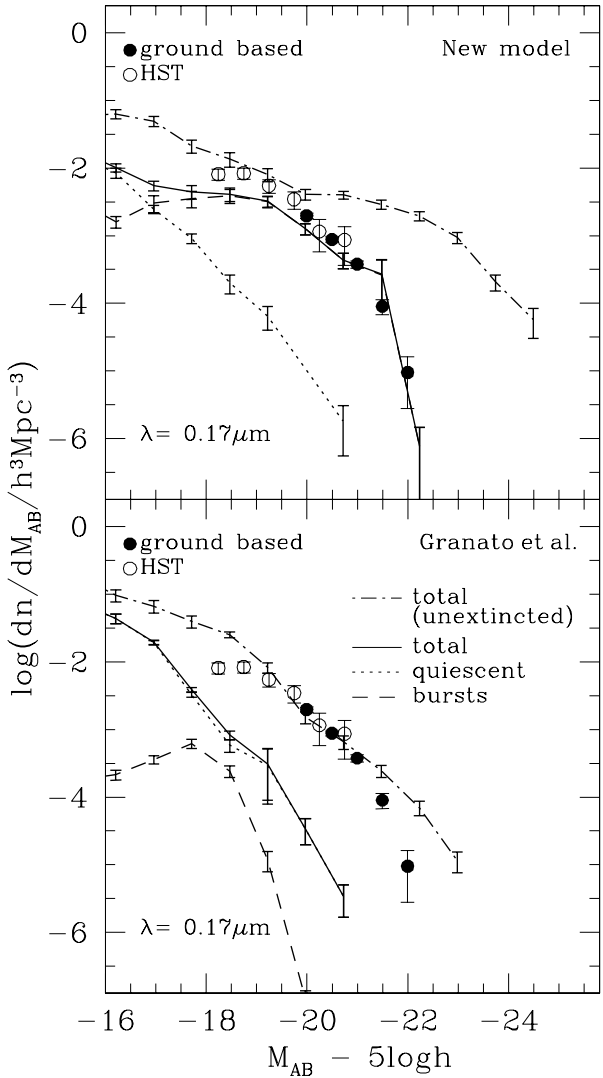}}
\caption{ The luminosity function in the rest-frame UV at $z=3$. The
top and bottom panels respectively show the predictions from the new
model presented in this paper and that of G00. In both panels, the
symbols show the observational estimate of the rest-frame UV
luminosity function of LBGs at $z\approx 3$ obtained by Steidel \etal
(1999) by combining ground-based and HST data. The rest-frame
wavelength of $0.17\mum$ corresponds to the observer-frame $\cal
R$-band used by Steidel \etal for $z=3$. In each panel, the solid line
shows the predicted total luminosity function including dust
extinction, while the dotted and dashed lines show the contributions
to this from quiescent galaxies and ongoing bursts respectively. The
dot-dashed line shows the total luminosity function without dust
extinction.  }
\label{fig:lfuv}
\end{figure}

\begin{figure}
{\epsfxsize=8.5truecm
\epsfbox[0 150 570 705]{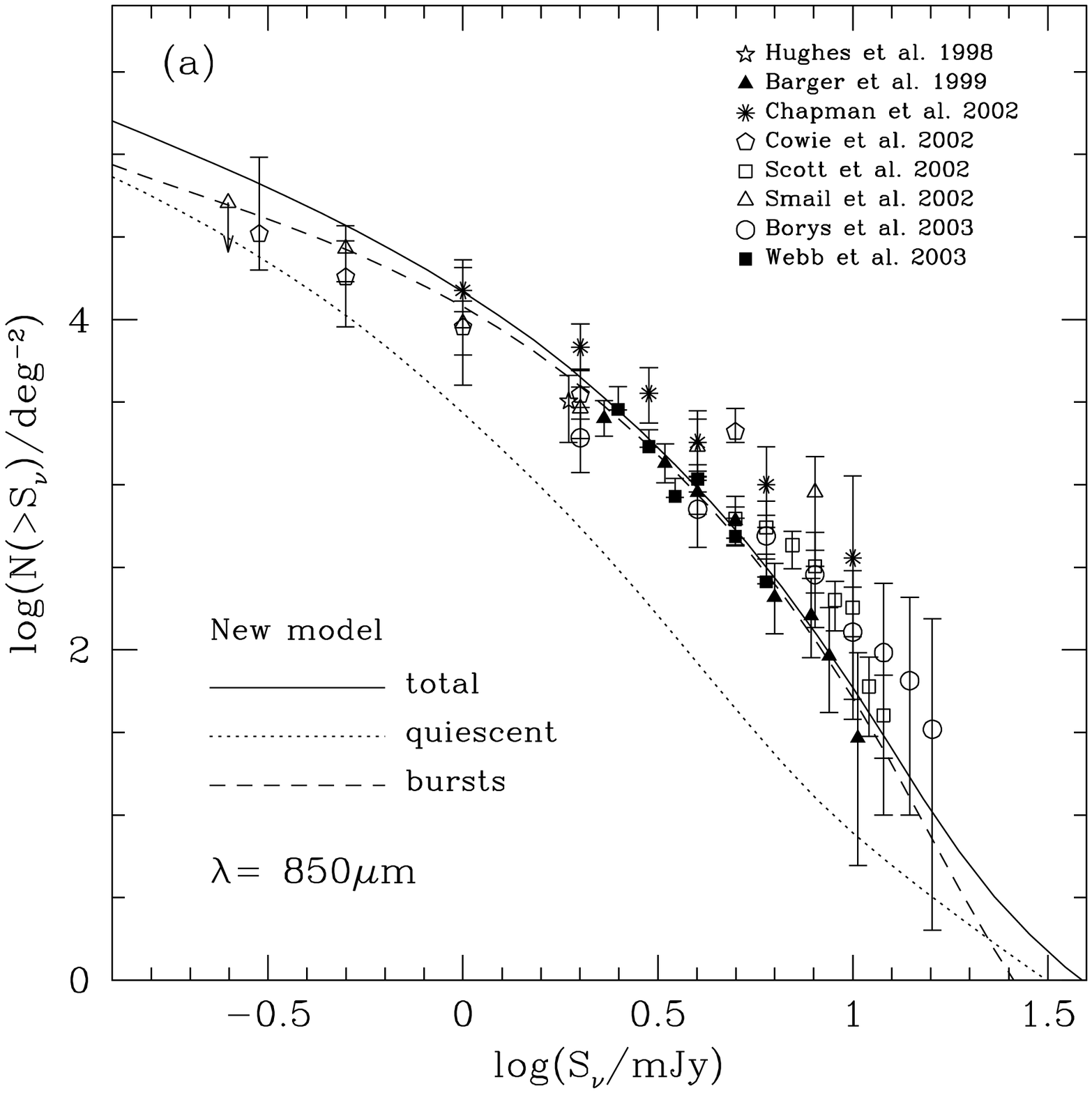}}
{\epsfxsize=8.5truecm
\epsfbox[0 150 570 705]{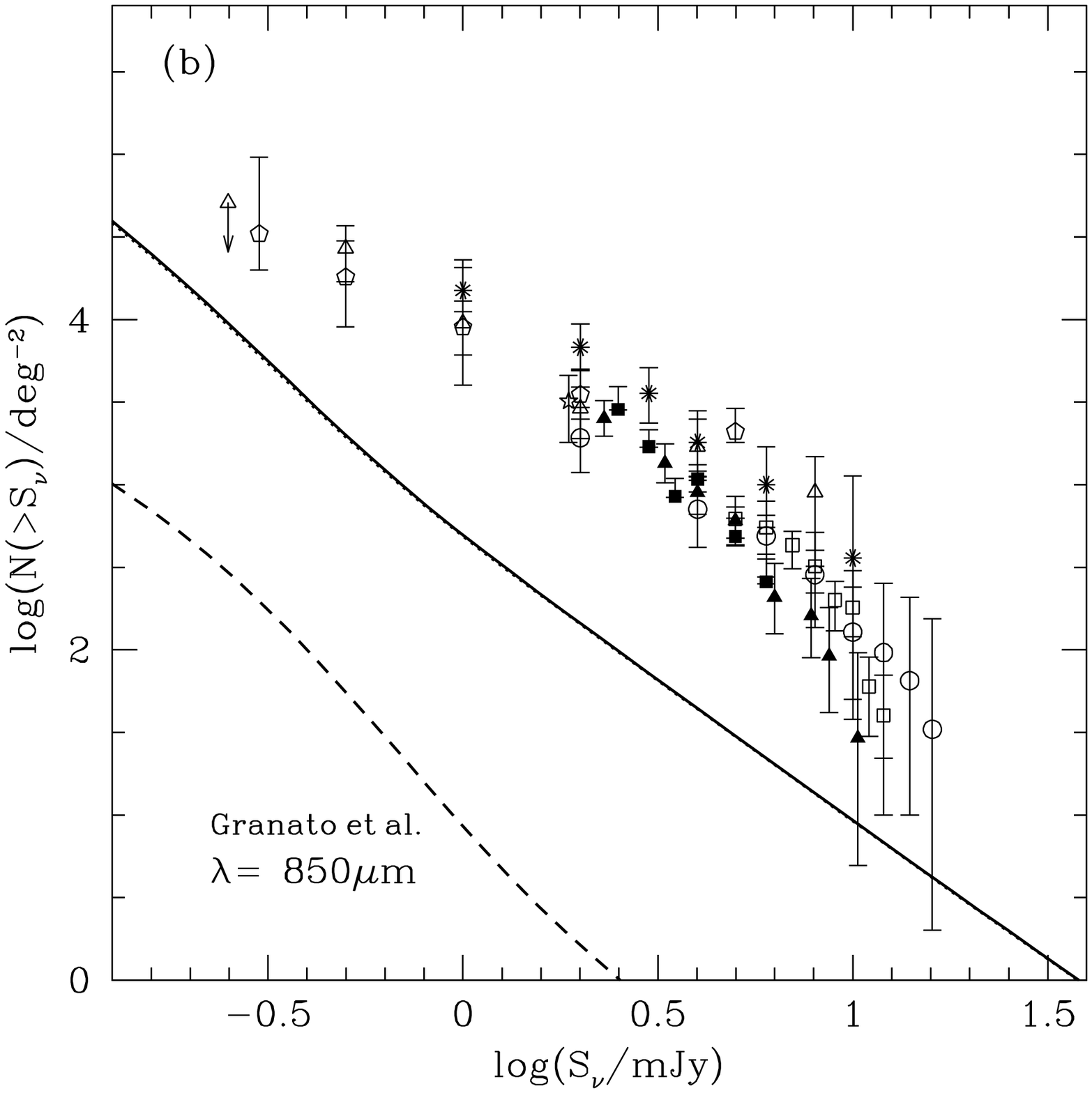}}
\caption{ The cumulative number counts per square degree at $850\mu
m$. The model predictions are shown by the lines: the total counts are
shown by the solid line, and the contributions from bursts and
quiescent galaxies by the short-dashed and dotted lines respectively.
(a) Shows the counts predicted by the new model and (b) shows the
predictions of the G00 model.  The symbols show a compilation of
observational data, as given in the key. }
\label{fig:n850}
\end{figure}

We first make some general comparisons between the predictions of the 
G00 model and our new model, before focusing on the properties of 
galaxies at high redshift.

{\em Cosmic star formation history:} We compare the global star
formation histories in G00 and the new model in
Fig.~\ref{fig:sfrv}. The overall star formation rate in the new model
(Fig.~\ref{fig:sfrv}a) is similar to that in G00
(Fig.~\ref{fig:sfrv}b).  There is a broad peak in the total star
formation density at $z \approx 2-3$, corresponding to a look-back
time of $\sim 10$Gyr.  However, in contrast to G00, where star
formation in bursts is around 5 per cent of the total at all
redshifts, in the new model the fraction of star formation occuring in
bursts increases sharply with redshift, and exceeds that in
quiescent disks for $z\gsim 3$.  This increase in the importance of
bursts at high-z relative to G00 results both from the change in the
star formation timescale in quiescent disks, which causes them to be
more gas-rich at high z, and from allowing bursts to be triggered by
minor mergers.  The fraction of star formation in bursts summed over
the history of the universe remains modest, however.  Only 30 per cent
of star formation takes place in the burst mode in the new model. The
contribution of bursts to the mass locked up in stars is even less
impressive. Taking into account the large fraction of mass recycled by
dying stars with the top-heavy IMF, less than 7 per cent of the mass
locked up in stars today is predicted to have been made in bursts.

{\em Present-day galaxy luminosity function:} The present-day galaxy
luminosity function plays an essential role in constraining the
parameters of the galaxy formation model.  Our philosophy is to
consider only those models that produce reasonable matches to the
optical galaxy luminosity function, tracing the stellar population,
and the far-IR luminosity function, tracing the emission from
dust. 
Fig.~\ref{fig:lfz0} compares the predictions of the new model with the
observed luminosity functions at $z=0$ in the optical (B-band) and in
the far-IR ($60\mum$).  In Fig.~\ref{fig:lfz0} (and
Fig. \ref{fig:lfuv}), the error bars on the model curves represent the
statistical uncertainties resulting from the finite sample of galaxies
computed; they are calculated as described in G00.  The new model
reproduces the local optical and far-IR luminosity function data
almost as well as the model of G00.  As in G00, ongoing bursts
dominate the bright end of the $60\mu$m luminosity function but are
less important in the optical. 
The new model model also reproduces other properties of the 
local galaxy population, such as the ratio of gas mass to luminosity as 
a function of luminosity and the stellar metallicity versus luminosity 
relation (see figures 9 and 10 from Cole \etal 2000, which are very similar 
to the results from our new model).
Further comparisons of the new model
with observed properties of present-day galaxies will be given in
Lacey \etal (2004a).

{\em Lyman-break galaxies:} A very important prediction of the model
is the galaxy luminosity function in the rest-frame UV at high
redshift. This depends both on the distribution of galactic star
formation rates and on the amount of extinction by dust.  We plot the
rest-frame UV luminosity function at $z=3$ in Fig.~\ref{fig:lfuv},
comparing the model predictions with observational data for LBGs. The
top panel shows the predictions of our new model, while the lower
panel shows the fiducial model of G00. The figure shows that the G00
model would match the observed UV luminosity function if there were no
dust extinction, but once the effects of dust have been included in a
consistent way, the G00 model is typically 2 magnitudes too faint at a
given galaxy abundance. In the new model, the dust-exticnted 
UV luminosity function matches the observations well. 
The luminosity function would be brighter by 1.5-2.5 magnitudes  or 
a factor 4--10 in luminosity in the absence of dust.
The level of extinction in our model galaxies 
is similar to that inferred by Steidel \etal (1999) using a  
simple extinction law and the observed $G-{\sc R}$ colours of their 
Lyman-break sample. However, a detailed comparison requires reproducing
the observational LBG selection in our model, and we defer this to a
future paper (Lacey \etal 2004b).

Since, on average, only a small fraction of
the UV light escapes from galaxies with large intrinsic UV
luminosities, the dust-extincted UV luminosity function is sensitive
to the parameters of the dust model. In the new model, we have adopted
a value for the escape time of stars from molecular clouds of
$\tesc=1\Myr$ for both bursts and quiescent star formation in order to
match the observed luminosity function of LBGs.  For this choice,
ongoing bursts dominate the 
\begin{figure*}
\begin{picture}(400,500)
\put(-50,250)
{\epsfxsize=8.5truecm
\epsfbox[20 150 570 705]{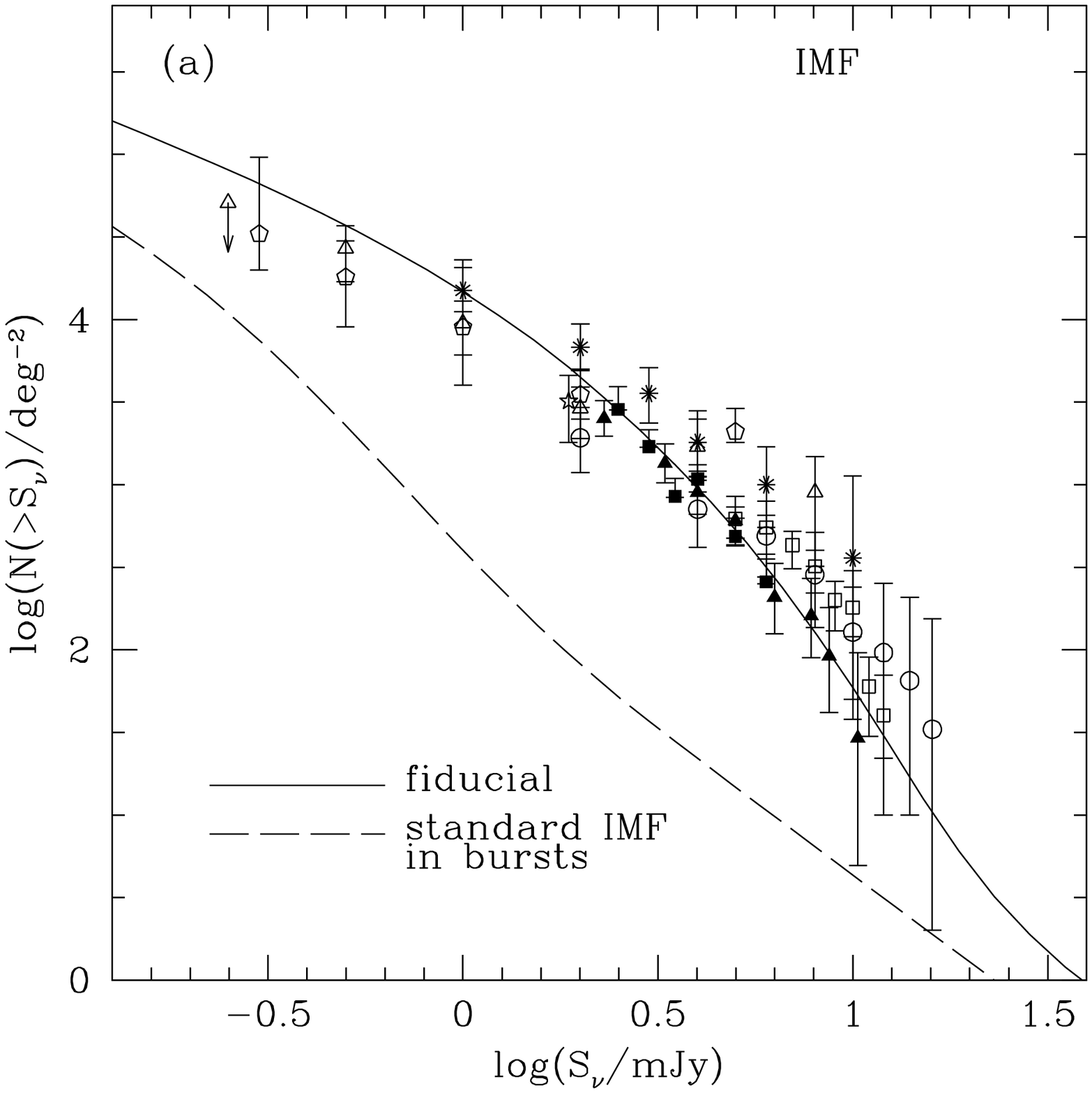}}
\put(200,250)
{\epsfxsize=8.5truecm
\epsfbox[20 150 570 705]{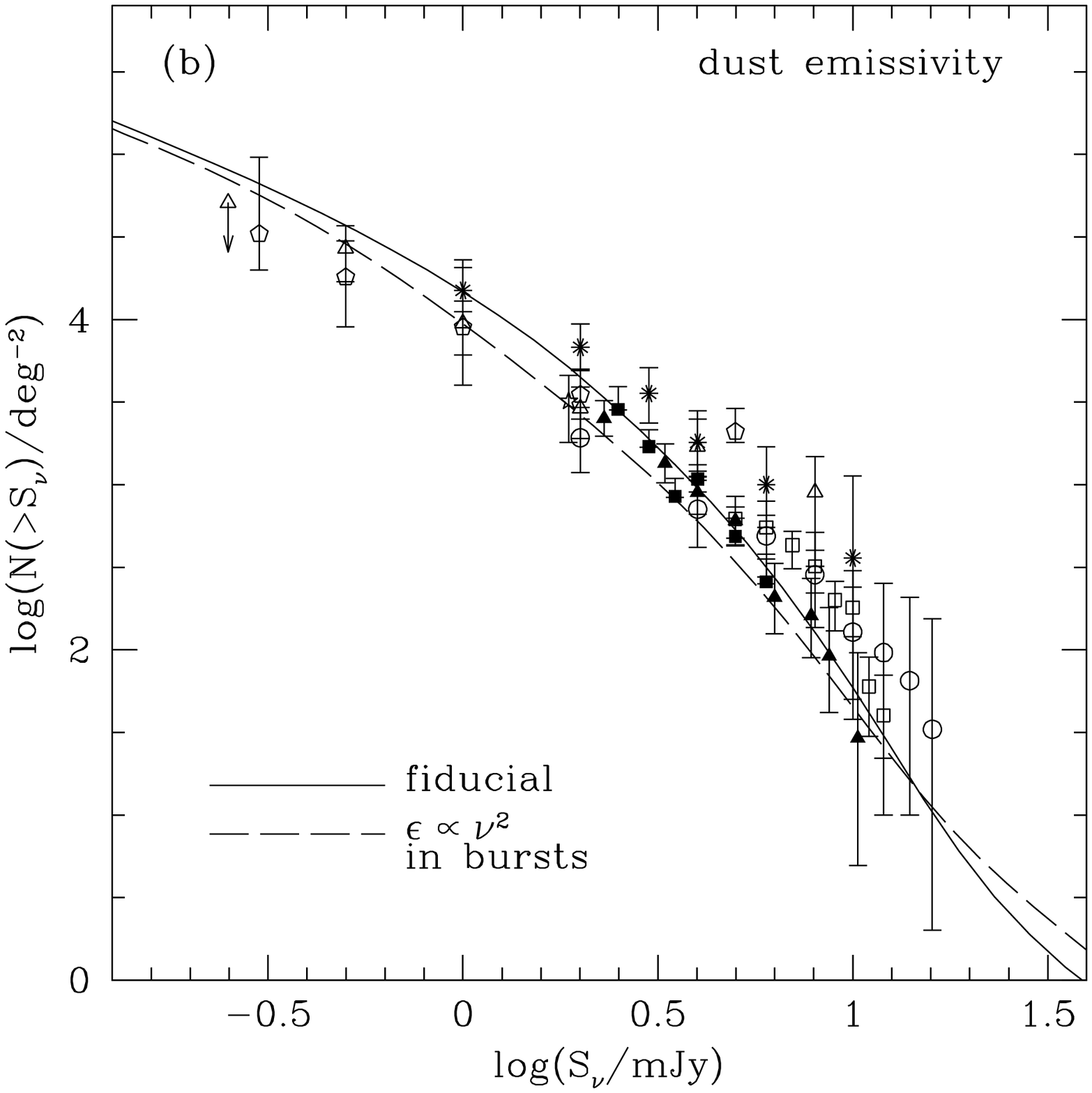}}
\put(-50,0)
{\epsfxsize=8.5truecm
\epsfbox[20 150 570 705]{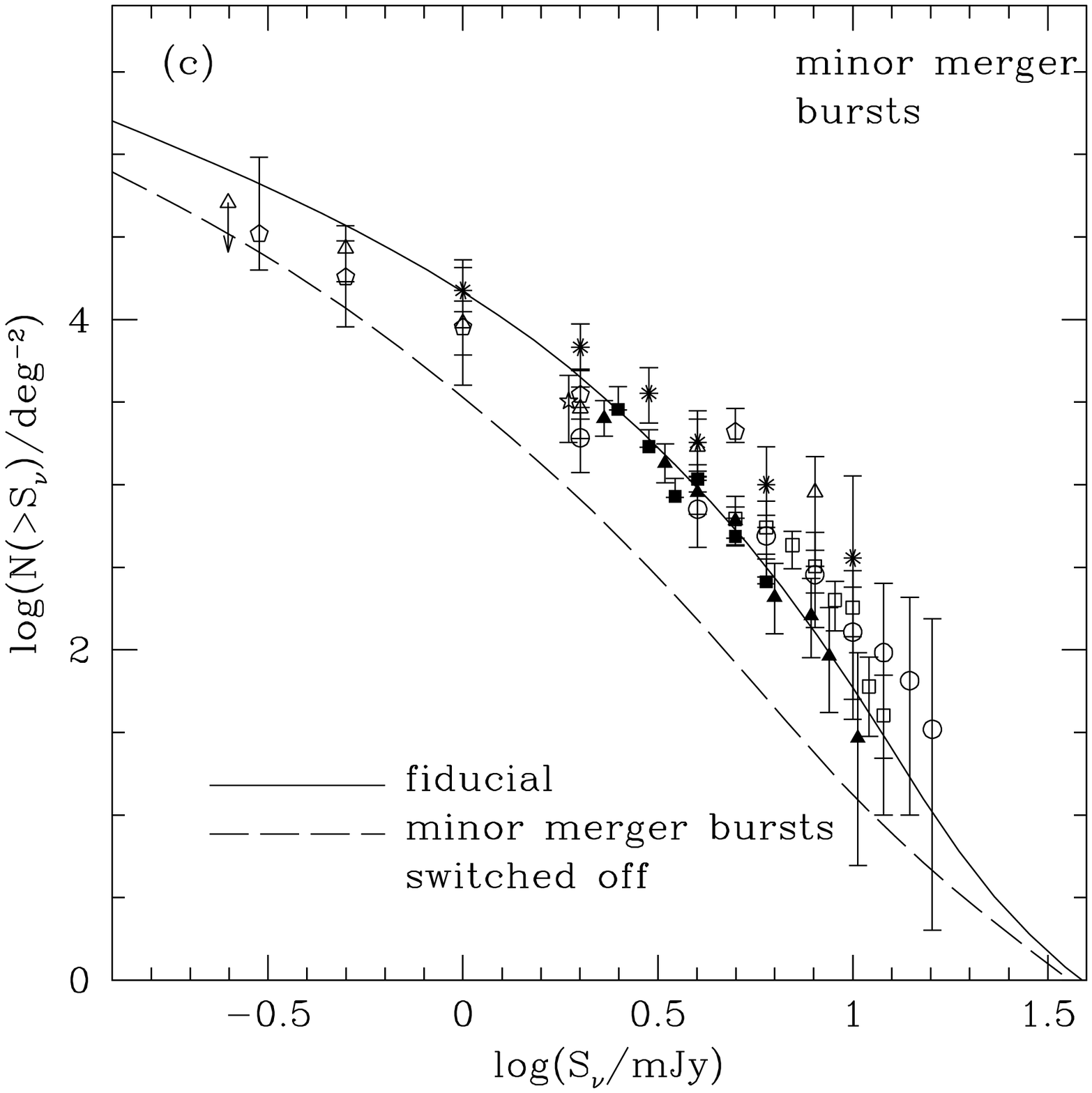}}
\put(200,0)
{\epsfxsize=8.5truecm
\epsfbox[20 150 570 705]{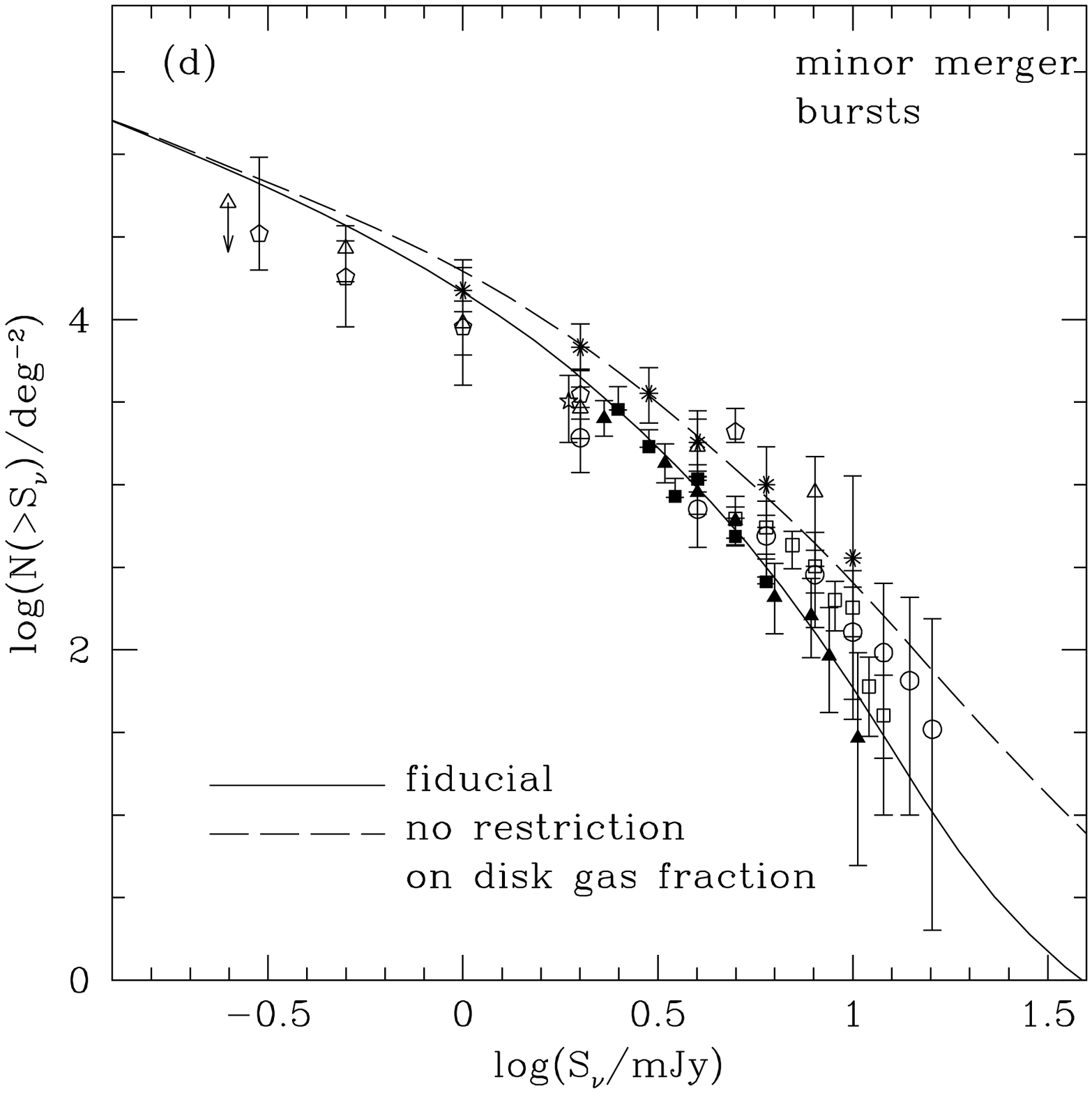}}
\end{picture}
\caption{ The effect of varying key model parameters and assumptions
on the predicted cumulative number counts per square degree at
$850\mum$.  The total source counts predicted by the new fiducial
model are shown by the solid line reproduced in each panel.  The
dashed lines show the predicted total counts in variant models: (a)
when a standard solar neighbourhood IMF is used in
bursts of star formation, (b) when the emissivity of
the dust in bursts matches that in quiescent galaxies, (c) when 
bursts triggered by minor mergers are switched off and (d) when the 
condition that the primary galaxy be gas rich in order to trigger a 
burst in a minor is switched off. 
The observational data are the same as those plotted in Fig.~\ref{fig:n850}.
}
\label{fig:changes}
\end{figure*}
extincted UV luminosity function at high
z.  However, our current choice of dust parameters may not be the only
one that allows a match to the observational data.  The
sub-mm emission from these galaxies is much less sensitive 
than the UV emission to the
details of the dust model, such as cloud masses, radii and escape
times, provided that most of the UV light is absorbed.  
The impact of varying key assumptions of the new model on the abundance 
of Lyman-break galaxies will be explored in a forthcoming paper, 
in which we will also explore in detail the connection between the 
sub-mm fluxes and rest-frame UV luminosities of high-z galaxies, 
and between SMGs and LBGs (Lacey \etal 2004b).

{\em Sub-mm galaxies:} The other crucial constraint on high-redshift
galaxies comes from the number counts of faint
SMGs. Fig.~\ref{fig:n850} shows the cumulative number counts as a
function of flux for galaxies selected by their emission at
$850\mum$. The model predictions are compared to a compilation of data
obtained by different surveys using SCUBA. The model of G00, shown in
Fig.~\ref{fig:n850}b, is seen to underpredict the observed counts, by
a factor $\sim 20$ at 3mJy. The new model presented in this paper is
much more successful (Fig.~\ref{fig:n850}a), agreeing spectacularly
well with the observations.  To obtain a better fit to the number
counts of SMGs, we have modified the emissivity of dust grains in
bursts at wavelengths $\lambda> 100\mum$, from the canonical
$\epsilon_{\nu} \propto \nu^{2}$ (which we retain for quiescent star
formation, and which G00 used for both modes of star formation), to
$\epsilon_{\nu} \propto \nu^{1.5}$.  There is independent evidence
that the dust emissivity in some starbursts may be flatter than
$\epsilon_{\nu} \propto \nu^{2}$: S98 found that modelling the SED of
Arp 220 favoured $\epsilon_{\nu} \propto \nu^{1.6}$.
As reviewed by Draine (2003), simple theoretical models of dust grains
predict that the dust emissivity should vary as $\epsilon_{\nu}
\propto \nu^{2}$ at long enough wavelengths, independently of the grain
size distribution. However, laboratory studies of some possible dust
materials find a much shallower dependence than this even at sub-mm
wavelengths. Assuming an emissivity like $\epsilon_{\nu} \propto
\nu^{1.5}$ thus remains controversial. We illustrate the impact on our
predictions of reverting to the canonical choice of $\epsilon_{\nu}
\propto \nu^{2}$ later on in this section.

\begin{figure}
{\epsfxsize=8.5truecm
\epsfbox[30 170 570 700]{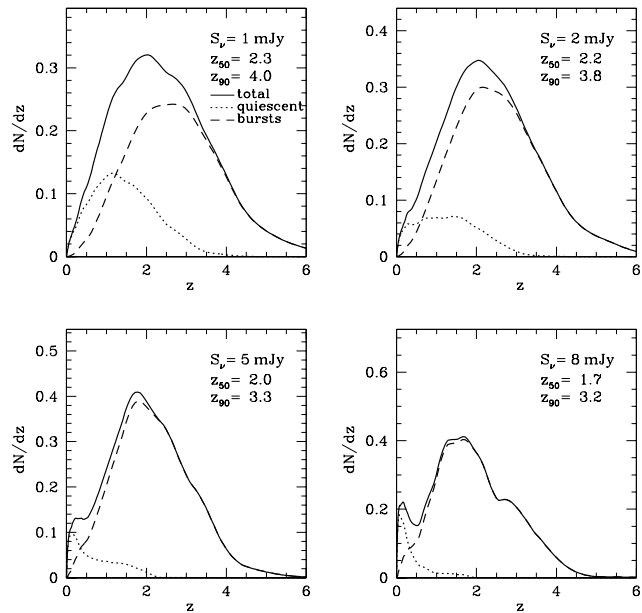}}
\caption{ The predicted redshift distributions of sources selected by
their $850\mum$ flux. Each panel corresponds to sources at a
particular flux, as indicated, increasing from $1\mjy$ at top left to
$8\mjy$ at bottom right. The solid line shows the redshift
distribution for all galaxies, and the dashed and dotted lines show
the contributions to this from ongoing bursts and quiescently star
forming galaxies respectively. The redshift distributions at each flux
are normalized to unit area for the total counts. The median redshift
$z_{50}$ and also the redshift $z_{90}$ below which $90\%$ of the
galaxies lie are given in each panel.}
\label{fig:dndz}
\end{figure}

In the new model, the counts are dominated by ongoing bursts at fluxes
$0.1 \lsim S_{\nu}(850\mum) \lsim 30\mjy$. Bursts account for a factor
of $10^3$ times more sources at $3\mjy$ in the new model than they did
in G00; the counts from sources that are not experiencing an ongoing
burst have increased by a more modest factor, $\approx 1.5$--$6$,
depending on the flux.

We now investigate the impact on the counts of SMGs of systematically
switching off, one at a time, the changes made in the new model. In
Fig.~\ref{fig:changes}, the total counts predicted by the new model
are shown by the solid line, which is reproduced in each panel. The
total counts in the variant models are shown by the dashed line in
each panel. Note that in this experiment, we have not altered any
additional parameters beyond the ones stated; however, in all cases,
the variant models predict optical luminosity functions at $z=0$ that
are in reasonably good agreement with the observational estimates.  In
Fig.~\ref{fig:changes}(a), we illustrate the effect of using a
standard Kennicutt IMF in both modes of star formation. This has the
biggest impact on the model predictions, reducing the counts by a
factor of $\approx 60$ at $3\mjy$.  The resulting total counts are similar
to those predicted by the G00 model. 
Changing the emissivity of the dust in bursts to be
the same as in the quiescent mode, $\epsilon_{\nu} \propto \nu^{2}$,
has a much less dramatic effect, reducing the counts by less than a
factor of 2 at $3\mjy$ (Fig.~\ref{fig:changes}b). Given the
observational scatter, this variant could be a viable alternative
model, especially if dust-obscured active galactic nuclei made a
comparable contribution to the $850\mum$ emission to that of dust-obscured
starbursts (although Alexander \etal 2003 argue that this is unlikely
to be the case).

The importance of bursts of star formation triggered
by minor mergers is illustrated in Fig.~\ref{fig:changes}(c), in which
we show how the predicted counts change when minor merger bursts are
turned off by setting $f_{\rm burst}=f_{\rm ellip}$. The counts at
$3\mjy$ are down by a factor of $\approx 7$ without the contribution
of minor merger bursts.  Finally, in Fig.~\ref{fig:changes}(d) we show
how the predicted counts are affected when we relax the condition for
triggering bursts in minor mergers that the gas fraction in the
primary galaxy should exceed $f_{\rm gas,crit}$, by setting $f_{\rm
gas,crit}=0$.  This variant model slightly overpredicts the $850\mum$
source counts.  However, this model greatly overpredicts the number of
bright $60\mum$ galaxies at the present day. As a direct consequence of
the parameterization of the star formation timescale (discussed in
Section \ref{ssec:advanced} above), galactic disks at high redshift
tend to be gas-rich in the new model, whereas they become
progressively more gas-poor as $z=0$ is approached, through the
depletion of gas by quiescent star formation.  The gas fraction
threshold therefore has the effect of suppressing most minor merger
bursts at low redshift, but has relatively little effect at high
redshift.

In Fig.~\ref{fig:dndz} we show what the new model predicts for the
redshift distributions of sources selected at $850\mum$. The median
redshift is seen to vary little over the flux range $1 <
S_{\nu}(850\mum) < 8\mjy$, from $z_{50}=2.3$ at the faint end to
$z_{50}=1.7$ at the bright end. At a flux of $5\mjy$, the median
redshift is predicted to be 2.0, in good agreement with the recent
estimate by Chapman \etal (2003, 2004) from a sample of SMGs with
spectroscopic redshifts. Fig.~\ref{fig:dndz} also shows that the
redshift distributions for the quiescent galaxies peak at lower
redshifts than the bursts. The quiescent galaxy peak is at very low
redshifts for the brighter part of the flux range shown. Low-z
quiescent galaxies are predicted to dominate the total counts at
fluxes $S_{\nu}(850\mum) \gsim 30\mjy$.

We will discuss the predictions of the new models for the properties
of the SMGs and LBGs at length in Lacey \etal (2004b). In summary, at
$z=2$, SMGs with $S_{850\mum}\ge5\mjy$ have median stellar masses of
$\sim 10^{10}h^{-1}M_{\odot}$ and reside in haloes with a
median mass of $\sim 10^{12}h^{-1}M_{\odot}$. The SMGs are thus
predicted to reside in the more massive haloes in place at $z=2$, and
therefore will be more strongly clustered than the dark matter at this
epoch (Baugh \etal 2004), consistent with tentative observational 
constraints (Blain \etal 2004).

\section{Discussion and Conclusions}
\label{s:conc}

We have combined two powerful theoretical techniques, semi-analytical
modelling of galaxy formation and radiative transfer modelling of the
reprocessing of stellar radiation by dust, to address the issue of how
faint sub-mm galaxies fit into current theories of galaxy
formation. Our approach to this problem represents a significant
advance over previous work in this area for two reasons.  Firstly, our
overriding philosophy is to require that any acceptable model must
reproduce a basic set of observational data for present-day galaxies,
including the optical, near-IR and far-IR luminosity functions, gas
fractions and metallicities, and the distribution of disk scale sizes.
Secondly, the dust extinction and emission are computed
self-consistently, rather than being put in by hand.  Our adopted
methodology severely restricts the range of available parameter space,
making a successful reproduction of the high redshift galaxy
population more challenging, but, at the same time resulting in a model
with more predictive power.

The new model presented in this paper extends the successes of the G00 
model to high redshift. Our model matches the measured rest-frame UV
luminosity function of Lyman-break galaxies at $z\approx3$ (Steidel
\etal 1999). At the same time, it reproduces the number counts of
$850\mum$ sources, and predicts redshifts for these sources in very
good agreement with recent measurements from Chapman \etal (2003, 2004). 
This is the first time that any realistic CDM-based galaxy formation
model has successfully matched the properties of {\em both} major
populations of star-forming galaxies at high redshift.
 
In order to achieve these successes while still remaining within the
framework of the hierarchical assembly of galaxies, we found it
necessary to enhance the importance of bursts at high
redshift, by: (i) modifying the quiescent star formation timescale in
galactic disks, in order to make mergers at high redshift more gas-rich; (ii)
allowing triggering of bursts by minor as well as major mergers; and
(iii) adopting a top-heavy IMF in bursts. 

Our new model shares some features with those of Guiderdoni \etal
(1998) and Devriendt \& Guiderdoni (2000). They also proposed that the
fraction of star formation occuring in bursts increased strongly with
redshift, and suggested that the IMF in bursts was top-heavy, in order to
explain the cosmic far-IR background and the sub-mm number
counts. However, in other respects, their model is very different from
ours. In particular, their model does not include galaxy mergers, so
the rate of starbursts as a function of redshift is put in by
hand. Also, they use empirical SEDs for the IR/sub-mm emission, rather
than calculating dust temperatures from radiative equilibrium.


The most controversial of our changes is the adoption of a top-heavy
IMF in bursts of star formation. 
For simplicity, we have chosen to use a flat IMF in starbursts. 
We have explored a range of tilted and truncated IMFs. A flat IMF slope 
is not essential  to match the number counts of SMGs; we found that reasonable 
matches to the counts could be obtained with slopes given by $x<0.35$. 
The critical point is that the relative proportion of high mass 
to low mass stars produced in a starburst should be greater than 
it is in quiescent star formation. 

We are by no means the first to suggest the abandonment of a universal IMF. 
Some theoretical calculations predict a form for the 
IMF that is biased towards more massive stars, compared 
with the Kennicutt (1983) IMF, under 
the conditions that exist in starbursts (e.g. Padoan \etal 1997; Larson 1998). 
However, the bulk of the support for a variation in the IMF 
is indirect, resulting from the challenge of matching observations 
of early-type galaxies and explaining the metal content of the hot 
gas in clusters (e.g. Renzini \etal 1993; Chiosi \etal 1998; 
Gibson \& Matteucci 1997; Thomas \etal 1999). 
The metal content of the intracluster medium and stellar absorption 
line strengths  in ellipticals can be more readily 
explained if significant star formation occured with an IMF which had
a higher fraction of massive stars compared to that in the solar neighbourhood.
This boost in the relative proportion of high mass stars can be
achieved either by 
tilting the IMF or by truncating it below a fixed stellar mass. 
Nagashima \etal (2004) have taken the new model described in this paper 
and incorporated a detailed treatment of chemical enrichment by Type~Ia 
and Type~II supernovae.  The new model reproduces the observed abundances 
of O, Mg, Si and Fe in the intracluster medium of X-ray clusters.  
Nagashima \etal find that reverting to a standard IMF in bursts leads to 
the model underpredicting the abundances of these elements by a factor
of 2--3.

Granato \etal (2004) have recently presented an alternative model
which can explain the sub-mm number counts without assuming either a
top-heavy IMF or a modified dust emissivity, based on spheroid
formation by monolithic collapse at high redshift. They posit that the
formation of very massive stellar spheroids occurs much earlier than
in the model presented here, through the joint effects of very strong
feedback from QSOs suppressing the cooling of gas in small halos at
high redshift, and a rate of gas cooling in high-mass halos that is
greatly enhanced by clumping effects. This model also reproduces the 
inferred evolution of the K-band luminosity function (e.g. Pozzetti \etal 
2003: Drory \etal 2004). However, unlike the model
presented here, the Granato \etal model does not include the effects
of either halo or galaxy mergers, and does not currently treat the
formation of galactic disks. The differences between these two models
highlight the need to develop a better physical understanding of the
processes of gas cooling and feedback, in order to understand whether
the assembly history of galactic spheroids was closer to the
``monolithic'' or ``hierarchical'' pictures.

The relative contributions to the $850\mum$ fluxes of SMGs from dust
heating by AGNs and by starbursts have been the subject of much recent
debate.  The first Chandra X-ray images of areas which had been mapped
in the sub-mm with SCUBA revealed little overlap between the X-ray and
sub-mm sources, suggesting that SMGs were not powered by obscured QSOs
(e.g. Almaini \etal 2003).  Deeper X-ray images have revealed a much
closer correlation between X-ray and sub-mm sources, with 30-50\% of
bright SMGs ($S_{850\mum}>5\mjy$) having X-ray-detected AGN
counterparts, and some others showing X-ray emission consistent with
star formation (Alexander \etal 2003).  However, despite the apparent
commonness of AGNs in SMGs, Alexander \etal argue, based on an
analysis of SED shapes, that even in the sub-mm sources which host
AGNs, the AGN contributes on average $<2\%$ of the $850\mum$ flux. The
SMGs thus appear to be powered by star formation, as assumed in our
model.

In our new model, star formation in bursts is very important at high
redshift; however, bursts are still responsible for only $30\%$ of the 
total star formation when integrated over all redshifts, compared to
$5\%$ in G00. Once the recycling of mass from dying stars is taken
into account, less than $7\%$ of the mass locked up in stars today is
predicted to have been produced in bursts.  The new model and the G00
model produce similar stellar mass fractions in disks and spheroids;
in the new model $57\%$ of the stellar mass is in disks and $43\%$ is
in spheroids.  This is in very good agreement with the observational
estimate by Benson, Frenk \& Sharples (2002a).  The most massive
galaxies in the model tend to be bulge dominated, in agreement with
observations (e.g. Kochanek \etal 2001, Nakamura \etal 2003).
However, typically only a few per cent of the present day stellar mass
in these galaxies was formed in bursts at $z \sim 2 $, around the peak
in the predicted redshift distribution of $850\mum$ sources.  In our
model, the overwhelming bulk of the stellar mass of ellipticals is
built up in disks and then rearranged into spheroids during mergers
(Baugh \etal 1996; Kauffmann 1996).

Our model is able to address many other issues relating to sub-mm
galaxies and to galaxy evolution generally, which will we pursue in
subsequent papers in this series. These include: a more detailed
comparison with galaxy properties in the local universe (Lacey \etal
2004a); the environments and clustering of SMGs and LBGs (Baugh \etal
2004); a more detailed analysis of the properties of LBGs and SMGs and
of the relationship between them (Lacey \etal 2004b); the descendants
of SMGs (Frenk \etal 2004); and a comprehensive set of predictions for
number counts at different wavelengths in the IR and sub-mm, and for
the extragalactic background light (Lacey \etal 2004c).

\section*{Acknowledgments}
CMB and AJB acknowledge receipt of Royal Society University Research
Fellowships.  This work was supported in part by a PPARC rolling grant
at Durham. We thank Ian Smail for his helpful comments and encouragement, 
and the referee for providing a detailed and helpful report. We also 
acknowledge useful comments from Rowena Malbon.

\end{document}